\documentclass{article}

\usepackage{PRIMEarxiv}

\usepackage[utf8]{inputenc} 
\usepackage[T1]{fontenc}    
\usepackage{hyperref}       
\usepackage{url}   
\usepackage{acro}
\DeclareAcronym{ecc}{
short = ECC,
long = error correcting code
}
\DeclareAcronym{ber}{
short = BER,
long  = bit error rate
}
\DeclareAcronym{asic}{
short        = ASIC,
long         = application-specific integrated-circuit
}
\DeclareAcronym{fpga}{
short = FPGA,
long  = field-programmable gate array
}
\DeclareAcronym{gpu}{
short = GPU,
long  = graphics processing unit
}
\DeclareAcronym{ldpc}{
short = LDPC,
long = low-density parity-check
}
\DeclareAcronym{nbldpc}{
short = NB-LDPC,
long = non-binary low-density parity-check
}
\DeclareAcronym{dvbs2}{
short = DVB-S2,
long = ETSI digital video broadcasting \nth{2} generation
}
\DeclareAcronym{cn}{
short = CN,
long = check node
}
\DeclareAcronym{vn}{
short = VN,
long = variable node
}
\DeclareAcronym{spa}{
short = SPA,
long = sum-product algorithm
}
\DeclareAcronym{lspa}{
short = LSPA,
long = log sum-product algorithm
}
\DeclareAcronym{emsa}{
short = EMSA,
long = extended min-sum algorithm
}
\DeclareAcronym{mma}{
short = MMA,
long = min-max algorithm
}
\DeclareAcronym{pcm}{
short = PCM,
long = parity-check matrix
}
\DeclareAcronym{bn}{
short = BN,
long = bit node
}
\DeclareAcronym{qcldpc}{
short = QC-LDPC,
long = quasi-cyclic low-density parity-check
}
\DeclareAcronym{gf}{
short = GF,
long = Galois field
}
\DeclareAcronym{mlgd}{
short = MLDG,
long =  majority-logic decoding
}
\DeclareAcronym{ihrb}{
short = IHRB,
long =  iterative hard reliability-based
}
\DeclareAcronym{isrb}{
short = ISRB,
long =  iterative soft reliability-based
}
\DeclareAcronym{gbfda}{
short = GBFDA,
long =  generalized bit-flipping decoding algorithm
}
\DeclareAcronym{amsa}{
short = AMSA,
long =  Adaptive Multiset Stochastic Algorithm
}
\DeclareAcronym{wbrb}{
short = WBRB,
long =  weighted bit reliability-based
}
\DeclareAcronym{fbrb}{
short = FBRB,
long =  full bit reliability-based
}
\DeclareAcronym{mimo}{
short = MIMO,
long =  multiple-input multiple-output
}
\DeclareAcronym{qam}{
short = QAM,
long =  quadrature amplitude modulation
}
\DeclareAcronym{snr}{
short = SNR,
long =  signal-to-noise ratio
}
\DeclareAcronym{adbp}{
short = ADBP,
long =  analog digital belief propagation
}
\DeclareAcronym{srb}{
short = SRB,
long =  symbol reliability based
}
\DeclareAcronym{gps}{
short = GPS,
long =  global positioning system
}
\DeclareAcronym{vlsi}{
short = VLSI,
long =  very large scale integration
}
\DeclareAcronym{fht}{
short = FHT,
long =  fast Hadamard transform
}
\DeclareAcronym{fft}{
short = FFT,
long =  fast Fourier transform
}
\DeclareAcronym{hls}{
short = HLS,
long =  high-level synthesis
}
\DeclareAcronym{rtl}{
short = RTL,
long =  register transfer level
}

\DeclareAcronym{qos}{
short = QoS,
long =  quality of service
}

\DeclareAcronym{bp}{
short = BP,
long =  belief propagation
}

\DeclareAcronym{llr}{
short = LLR,
long =  log-likelihood ratio
}

\DeclareAcronym{csr}{
short = CSR,
long =  compressed sparse row
}

\DeclareAcronym{csc}{
short = CSC,
long =  compressed sparse column
}

\DeclareAcronym{lut}{
short = LUT,
long =  look-up table
}

\DeclareAcronym{cpu}{
short = CPU,
long  = central processing unit
}

\DeclareAcronym{sm}{
short = SM,
long  = streaming multiprocessor
}

\DeclareAcronym{ram}{
short = RAM,
long  = random access memory
}

\DeclareAcronym{ccsds}{
short = CCSDS,
long  = Consultative Committee for Space Data Systems
}

\DeclareAcronym{bch}{
short = BCH,
long  = Bose–Chaudhuri–Hocquenghem
}

\DeclareAcronym{fwht}{
short = FWHT,
long  = fast Walsh-Hadamard transform
}

\DeclareAcronym{tdp}{
short = TDP,
long  = thermal design power
}

\DeclareAcronym{cuda}{
short = CUDA,
long  = compute unified device architecture
}

\DeclareAcronym{pim}{
short = PiM,
long  = processing-in-memory
}

\DeclareAcronym{pnm}{
short = PnM,
long  = processing-near-memory
}

\DeclareAcronym{pum}{
short = PuM,
long  = processing-using-memory
}

\DeclareAcronym{dpu}{
short = DPU,
long  = DRAM processing unit
}

\DeclareAcronym{ms}{
short = MS,
long = min-sum
}

\DeclareAcronym{wram}{
short = WRAM,
long = working SRAM
}

\DeclareAcronym{mram}{
short = MRAM,
long = main DRAM
}

\DeclareAcronym{sram}{
short = SRAM,
long =static random-access memory
}

\DeclareAcronym{dram}{
short = DRAM,
long =dynamic random-access memory
}

\DeclareAcronym{mm}{
short = MM,
long =min-max
}

\DeclareAcronym{alu}{
short = ALU,
long = arithmetic logic unit
}

\DeclareAcronym{bpsk}{
short = BPSK,
long = binary phase-shift keying
}

\DeclareAcronym{awgn}{
short = AWGN,
long = additive white Gaussian noise
}

\DeclareAcronym{bf}{
short = BF,
long = bit-flipping
}

\DeclareAcronym{cnp}{
short = CNP,
long = check node processing
}

\DeclareAcronym{vnp}{
short = VNP,
long = variable node processing
}

\DeclareAcronym{app}{
short = APP,
long = \textit{a posteriori} probability
}

\DeclareAcronym{isa}{
short = ISA,
long = instruction set architecture
}

\DeclareAcronym{risc}{
short = RISC,
long = reduced instruction set computer
}

\DeclareAcronym{iram}{
short = IRAM,
long = instruction DRAM
}

\DeclareAcronym{fp}{
short = FP,
long = floating-point
}

\DeclareAcronym{dma}{
short = DMA,
long = direct memory access
}

\DeclareAcronym{dimm}{
short = DIMM,
long = dual in-line memory module
}

\DeclareAcronym{ddr4}{
short = DDR4,
long = double data rate 4
}

\DeclareAcronym{spmd}{
short = SPMD,
long = {single program, multiple data}
}

\DeclareAcronym{sdk}{
short = SDK,
long = software development kit
}

\DeclareAcronym{simd}{
short = SIMD,
long = {single instruction, multiple data}
}

\DeclareAcronym{simt}{
short = SIMT,
long = {single instruction, multiple thread}
}

\DeclareAcronym{raw}{
short = RAW,
long = {read-after-write}
}

\DeclareAcronym{ran}{
short = RAN,
long = {radio access network}
}

\DeclareAcronym{ann}{
short = ANN,
long = {artificial neural network}
}

\DeclareAcronym{mlp}{
short = MLP,
long = {multilayer perceptron}
}

\DeclareAcronym{nn}{
short = NN,
long = {neural network}
}

\DeclareAcronym{gemm}{
short = GEMM,
long = {general matrix multiplication}
}

\DeclareAcronym{mvm}{
short = MVM,
long = {matrix-vector multiplication}
}

\DeclareAcronym{mac}{
short = MAC,
long = {multiply-accumulate}
}

\DeclareAcronym{ai}{
short = AI,
long = {artificial intelligence}
}

\DeclareAcronym{dnn}{
short = DNN,
long = {deep neural network}
}

\DeclareAcronym{llm}{
short = LLM,
long = {large language model}
}

\DeclareAcronym{cnn}{
short = CNN,
long = {convolutional neural network}
}

\DeclareAcronym{fp32}{
short = FP32,
long = {32-bit floating point}
}

\DeclareAcronym{fp16}{
short = FP16,
long = {16-bit floating point}
}

\DeclareAcronym{tops}{
short = TOPS,
long = {tera operations per second}
}

\DeclareAcronym{mad}{
short = MAD,
long = {multiply-add}
}

\DeclareAcronym{hmc}{
short = HMC,
long = {Hybrid Memory Cube}
}

\DeclareAcronym{nvm}{
short = NVM,
long = {non-volatile memory}
}

\DeclareAcronym{adc}{
short = ADC,
long = {analog-to-digital converter}
}

\DeclareAcronym{dac}{
short = DAC,
long = {digital-to-digital converter}
}

\DeclareAcronym{sotmram}{
short = SOT-MRAM,
long = {spin orbit torque magnetic random access memory}
}

\DeclareAcronym{bnn}{
short = BNN,
long = {binary neural network}
}

\DeclareAcronym{reram}{
short = ReRAM,
long = {resistive random access memory}
}

\DeclareAcronym{hrl}{
short = HRL,
long = {heterogeneous reconfigureable logic}
}

\DeclareAcronym{cgra}{
short = CGRA,
long = {coarse-grained reconfigurable architecture}
}

\DeclareAcronym{dna}{
short = DNA,
long = {deoxyribonucleic acid}
}
\usepackage{booktabs}       
\usepackage{amsmath,amssymb,amsfonts}
\usepackage{nicefrac}       
\usepackage{microtype}      
\usepackage{lipsum}
\usepackage{fancyhdr}       
\usepackage{graphicx}       
\graphicspath{{media/}}     
\usepackage{algorithmic}
\usepackage{algorithm}
\usepackage{comment}
\usepackage{multirow}
\usepackage{xcolor}
\usepackage{soul}
\usepackage{bm}
\usepackage{tcolorbox}
\usepackage{siunitx}
\usepackage{subcaption}

\title{An Experimental Exploration of In-Memory Computing for Multi-Layer Perceptrons
}

\author{
  Pedro Carrinho$^{1\dagger}$, Hamid Moghadaspour$^{2,3\dagger}$, Oscar Ferraz$^{2,3\dagger}$, João Dinis Ferreira$^{2}$, Yann Falevoz$^{4}$,\\ \textbf{Vitor Silva$^{2,3}$, and Gabriel Falcao$^{2,3}$}\\
 $^{1}$Technology Engineering group, International Iberian Nanotechnology
Laboratory, Braga, Portugal\\
$^{2}$Instituto de Telecomunicações, Portugal\\
$^{3}$Department of Electrical and Computer Engineering, University of
Coimbra, Coimbra, Portugal\\
$^{4}$UPMEM, WTC Chambre Commerce Industrie, Grenoble, France\\
  \texttt{pedro.carrinho@inl.int; \{hamid.moghadaspour, oscar.ferraz, vitor, gff\}@co.it.pt};\\ \texttt{joaodinis.sf@gmail.com; yfalevoz@upmem.com} \\ \\
$^{\dagger}$These authors contributed equally to this work
}

\begin{document}
\maketitle

\begin{abstract}
In modern computer architectures, the performance of many memory-bound workloads (e.g., machine learning, graph processing, databases) is limited by the data movement bottleneck that emerges when transferring large amounts of data between the main memory and the central processing unit (CPU).
Processing-in-memory is an emerging computing paradigm that aims to alleviate this data movement bottleneck by performing computation \textit{close to} or \textit{within} the memory units, where data resides.

One example of a prevalent workload whose performance is bound by the data movement bottleneck is the training and inference process of artificial neural networks. 
In this work, we analyze the potential of modern general-purpose PiM architectures to accelerate neural networks.
To this end, we selected the UPMEM PiM system, the first commercially available real-world general-purpose PiM architecture.
We compared the implementation of multilayer perceptrons (MLPs) in PiM with a sequential baseline running on an Intel Xeon CPU.
The UPMEM implementation achieves up to $259\times$ better performance for inference of large batch sizes when compared against the CPU that exploits the size of the available PiM memory.
Additionally, two smaller MLP were implemented using UPMEM’s working SRAM (WRAM), a scratchpad memory, to evaluate their performance against a low-power Nvidia Jetson graphics processing unit (GPU), providing further insights into the efficiency of UPMEM's PiM for neural network inference. Results show that using WRAM achieves kernel execution times for MLP inference of under $3$ ms, which is within the same order of magnitude as low-power GPUs.
\end{abstract}


\section{Introduction}
\label{sec1}
The volume of digital data that is created, captured, copied, and consumed continues to increase rapidly.
Current projections predict it to surpass $394$ zettabytes by $2028$~\cite{big_data}.
In addition, many of the workloads operating on these large amounts of data (e.g., machine learning, graph processing, databases) are memory-bound in modern computer architectures~\cite{pim_workload_driven_prespective}.
Because of this, these architectures have struggled to keep up with end users' performance and energy efficiency expectations for these types of workloads.
Prior works identify the data movement bottleneck between the memory and the processing units as the key contributor to the observed performance bottleneck~\cite{pim_workload_driven_prespective,processing_data_where_it_makes_sense,modern_primer}.

A growing body of work proposes a paradigm shift from processor-centric to data-centric architectures, which can be achieved by processing the data close to or within the memory units, where it resides.
This processing paradigm, dubbed \textit{\acf{pim}}, has recently gained popularity as a way to mitigate the data movement bottleneck by enhancing memory chips with processing units.
This reduces the need for data to traverse the entire memory hierarchy for workloads that are memory-bound, providing substantial performance and energy efficiency benefits~\cite{modern_primer, processing_data_where_it_makes_sense, falevoz2023energy}.

In particular, \acp{nn} and other deep learning workloads are prevalent workloads that are often memory-bound in modern computer architectures.
This is primarily due to the \ac{gemm} and \ac{mvm} operations, which translate to \ac{mac} operations~\cite{pim_workload_driven_prespective,google_consuptions}.
These operations are often used to process large matrices that could greatly benefit from \ac{pim}, allowing for greater memory bandwidth and energy savings.

The UPMEM system~\cite{upmem_system} was the first general-purpose \ac{pim} system to be made available for both commercial and research use. It is composed of a host \ac{cpu}, standard main memory (\ac{dram} modules), and \ac{pim}-enabled memory (UPMEM memory modules). Each UPMEM module is composed of a \ac{dram} bank called \ac{mram}, accessible from the host \ac{cpu}, a scratchpad memory called \ac{wram}, and an instruction memory called \ac{iram}. In each \ac{pim} chip, there are 8 cores called \acp{dpu} which are able to access the three types of memory banks.


Previously in our study~\cite{carrinho2024processing}, we used the UPMEM system to explore the use of \ac{mram} to perform training and inference of larger \acp{nn}, and offer a comparison of the system with a \ac{cpu} baseline.
Since \ac{pim} is designed to alleviate the data movement bottleneck and reduce energy consumption, in this study we build upon our previous work by comparing the UPMEM system with a low-power \ac{gpu}, motivated by the increasing demand for efficient \ac{nn} inference, especially in edge computing and embedded \ac{ai} applications~\cite{singh2023edge}. Low-power \acp{gpu} are designed to provide massive parallelism while optimizing memory bandwidth usage, rendering them a fair comparison to \ac{pim} architecture \cite{desislavov2023trends}. Prior research indicates that memory access patterns and data locality significantly influence performance in both PIM-based architectures and low-power \acp{gpu}, especially when managing extensive neural networks and inference tasks~\cite{prim_other}. This comparison is crucial for comprehending how these architectures handle data transfer, memory bandwidth, and execution across environments.
Building upon our previous findings~\cite{carrinho2024processing}, this study makes the following \textbf{contributions}:
\begin{itemize}
    \item Comparison of smaller \acp{mlp} implemented on UPMEM's \ac{pim} system with a low-power \ac{gpu};
    \item A study on the viability of using not only \ac{mram} but also \ac{wram} for the inference of \acp{mlp}.
    \item Analyzing the data transfer to \ac{mram} and \ac{wram} and providing a comparison to low-power \acp{gpu}.
    
\end{itemize}

\section{Motivation}

\subsection{Neural Network Accelerators}
    
\Acp{nn} have become increasingly popular for tasks such as classification, detection, clustering, pattern recognition, and others, across many disciplines~\cite{sota_in_nn}. Accelerators play a crucial role by significantly reducing the time required for training and inference of large \acp{dnn}. With the growth of \ac{dnn} applications, such as \acp{llm}, there is a demand for energy-efficient hardware architectures~\cite{survey_accelerators}.
\Acp{nn} are memory-bound workloads~\cite{damov,google_consuptions} with the \ac{mac} operation representing the main arithmetic function used in \acp{nn}~\cite{survey_accelerators,damov,google_consuptions} and accelerators enable massive parallelism, allowing to satisfy the computational complexity which comes from the size increasing \acp{dnn}, with \acp{gpu} achieving speedups up to 90$\times$ for \acp{nn}~\cite{NN_survey_accelerators}.

\subsection{The Role of Processing-in-Memory in Neural Networks}

The emerging \ac{pim} computing paradigm is a promising alternative to address the data movement bottleneck by bypassing costly off-chip data transfers.
\ac{pim} represents a shift from a processing-centric paradigm towards a more data-centric paradigm~\cite{processing_data_where_it_makes_sense, modern_primer}, thus becoming a possible solution to overcome memory-bound operations, and mitigating the data movement bottleneck by making computations where the data resides, i.e., in the memory itself.

\ac{pim} tries to mitigate this bottleneck by either 1) enhancing the memory chips with processing elements in the memory itself (\ac{pum}~\cite{ambit2}), making the computation take place where the data reside, or 2) moving the processing units closer to the memory chips, bypassing their memory hierarchy (\ac{pnm})~\cite{processing_data_where_it_makes_sense, modern_primer}.

On average, $63\%$ of the energy spent on consumer device workloads come from data movement that could benefit from \ac{pim}~\cite{google_consuptions}.
In \acp{dnn}, data movement is more expensive than the computations~\cite{estimate_dnn_energy_consumption}.
For GoogLeNet, $10\%$ of the total energy is consumed on computations, whereas $68\%$ is spent on data movement~\cite{googlenet}.
The workloads studied in~\cite{google_consuptions} include the widely used deep learning framework, TensorFlow Lite, with the authors of~\cite{pim_workload_driven_prespective} concluding that for different \ac{cnn} models, during inference, up to $83\%$ of energy reduction is possible for packing and quantization operations using \ac{pim}.
These results came from the larger bandwidth and lower latency provided by the \ac{pim} device and having the \ac{cpu} executing the \ac{gemm} operation in parallel.

 
\subsection{Available PIM Systems for Neural Network Processing}

Many prior works describe \ac{pim} accelerators for \ac{nn}~\cite{cpim_hardware_accelerator_comparator, floatpim_accelerator, isaac_cnn_accelerator, prime_reram_accelerator, heterogeneous_approach_acceleration, zpim_accelerator_architecture, enna, puma, parapim, bit_precision_nn_reconfigurable_architecture, pnm_in_gpu}.
However, these accelerators rely on dedicated hardware.
Other works, such as Samsung AXDIMM~\cite{samsung_axdimm}, and Samsung HBM-PIM~\cite{samsung_hbm_pim} allow acceleration for \acp{nn}.
However, these \ac{pim} accelerators are not available to the general consumer, and the computing units do not behave like a traditional \ac{cpu} with a full \ac{isa}. 

In this work, we use UPMEM \ac{dram} chips for \ac{nn} training and inference, which offer fully-functional \acp{nn} in general-purpose \ac{pim} system in contrast to what is available in the literature.


General-purpose \ac{pim} accelerators are still in their infancy, being proposed and developed, thus, testing workloads on real-world scenarios is required to evaluate the merits of \ac{pim} as an accelerator for \acp{nn}.
For this reason, in \cite{carrinho2024processing} we created what is, to the best of our knowledge, the first \ac{nn} implementation to run in a commercially available \ac{pim} architecture. 



 \begin{figure}[t!]
\centering
\includegraphics[width=0.75\textwidth]{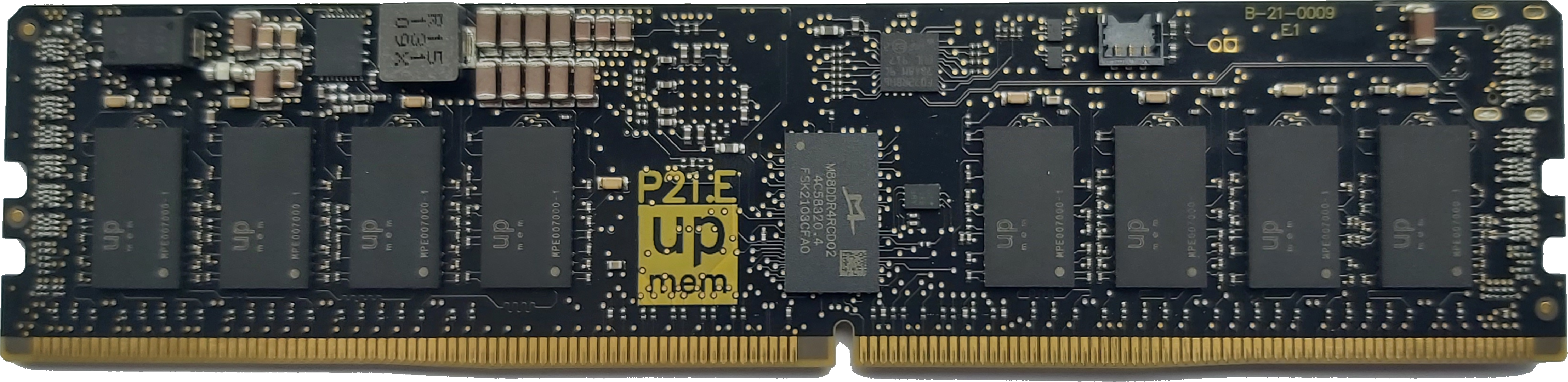}
\caption{UPMEM \ac{dram} \ac{dimm}, courtesy of~\cite{upmem_hotchips}.}
\label{fig:upmem_dimm}
\end{figure}


\section{UPMEM PIM Architecture}

\begin{figure}[t]
\centering
\includegraphics[width=\textwidth]{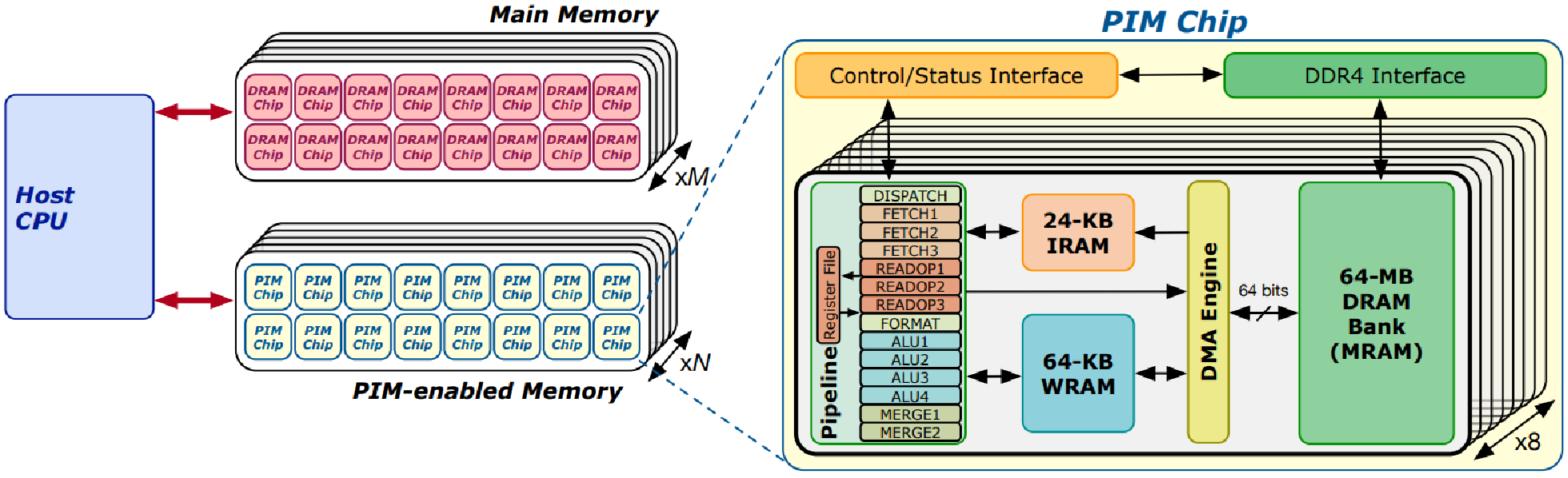}
\caption{UPMEM system architecture.}
\label{fig:PIM_DIMM_v1}
\vspace*{-10pt}
\end{figure}

The UPMEM system introduces a new DDR$4$-$2400$ \ac{dimm}, shown in Fig.~\ref{fig:upmem_dimm}, that can be plugged into a standard \ac{dimm} slot~\cite{upmem_hotchips}.
The \ac{cpu} handles the orchestration of the UPMEM system in a host-device setting.  
No intelligent memory controllers and cache coherence mechanisms are implemented, and it is up to the programmer to control the data to be sent to UPMEM memory.
In the UPMEM system, both main \acp{dimm} and UPMEM \ac{pim}-enabled \acp{dimm} co-exist in the same computing system and are controlled by the host \ac{cpu} as shown in Fig.~\ref{fig:PIM_DIMM_v1}.

\begin{figure}[t]
\centering
\includegraphics[width=0.7\textwidth]{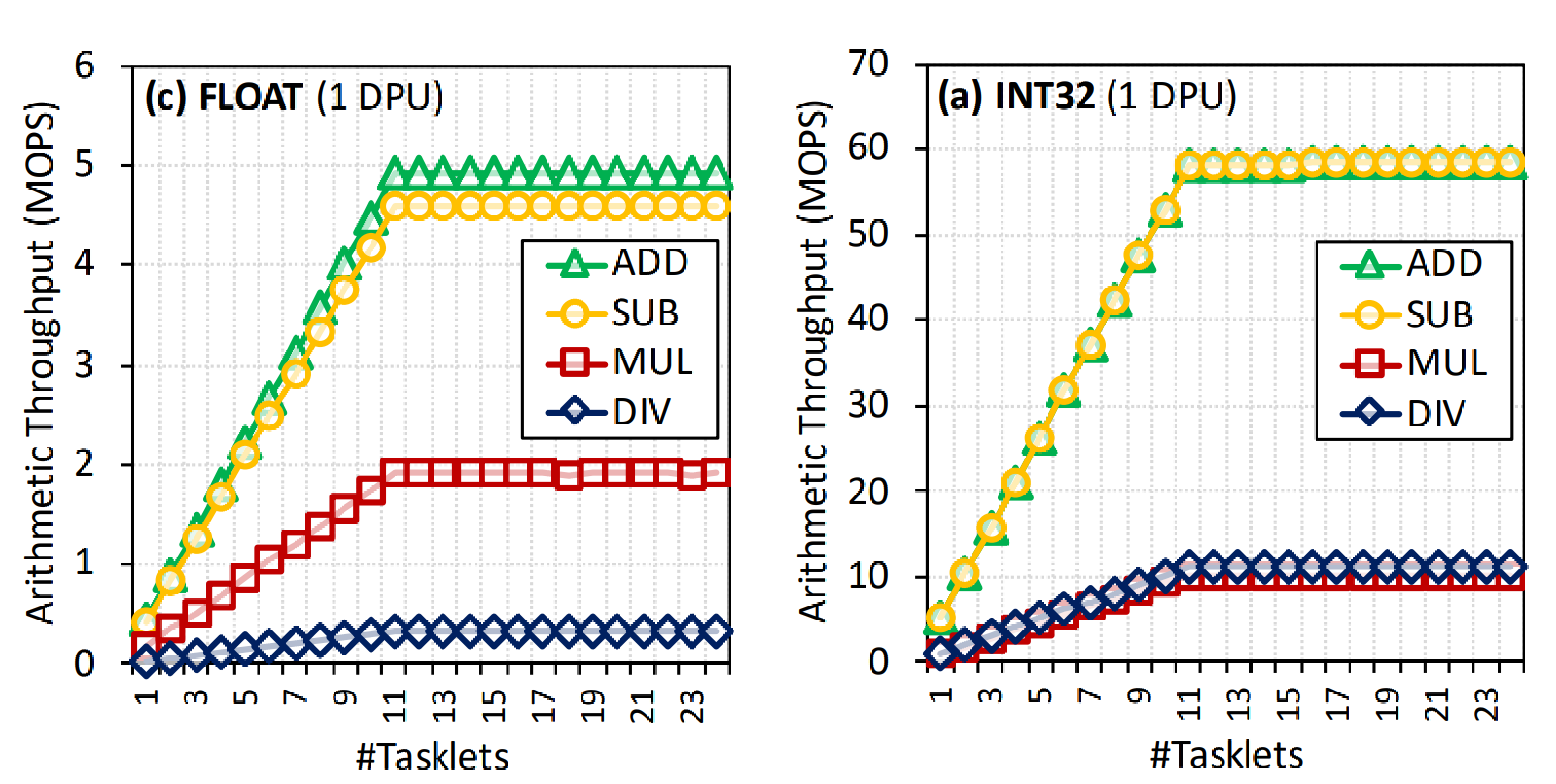}
\caption{Arithmetic throughput of four operations in \ac{fp32} and 32-bit integer for one \ac{dpu}, courtesy of~\cite{prim}.}
\label{fig:tasklets}
\end{figure}

A \ac{dpu} is a $32$-bit general-purpose processor with a dedicated \ac{isa} and $14$-stage pipeline.
The last three stages (ALU$4$, MERGE$1$, and MERGE$2$ of Fig.~\ref{fig:PIM_DIMM_v1}) can be executed in parallel with the DISPATCH and FETCH stages, thus taking $11$ cycles to perform an instruction and allowing $11$ threads to perform different stages in the same cycle.
Therefore, using more than $11$ threads does not improve performance~\cite{prim}, as depicted in Fig.~\ref{fig:tasklets}.
Instructions are sent to the \ac{dma} engine to move data between the $64$-MB \ac{mram} and the $64$-KB\ac{wram} and to move instructions from the \ac{mram} to the $24$-KB \ac{iram}, as shown in Fig. \ref{fig:PIM_DIMM_v1}.

The programmer determines the number of \acp{dpu} allocated, and the \acp{dpu} can execute in synchronous or asynchronous mode.
In synchronous mode, the host waits for all \acp{dpu} to finish, while in asynchronous mode, the \acp{dpu} perform concurrent execution and the host is free to perform other tasks.
Each \ac{dpu} executes up to $24$ software threads called tasklets, each capable of sharing resources inside the same \ac{dpu}.
Data from different \acp{dpu} can only be shared through the host. \ac{wram} memory is allocated inside the \ac{dpu} manually by the programmer~\cite{upmem_user_manual}.

The UPMEM architecture supports arithmetic and logic operations. However, for $16$-bit, $32$-bit, and $64$-bit floating-point multiplication and representation, it relies on software emulation, with only $8$-bit integer multiplication natively supported.

The commercially available UPMEM servers possess $20$ \ac{pim} \acp{dimm} totaling $2560$ \acp{dpu}, each running at $350$ MHz, and $160$ GB of \ac{pim} memory, providing up to $1.792$ TB/s of memory bandwidth, and four $64$ GB DDR$4$-$2666$ standard \ac{dram}~\cite{upmem_white_paper}.
The \acp{cpu} included in the system are based on Intel Xeon x$86$ ~\cite{upmem_system}.

\section{Multi-layer Perceptron}

The \ac{mlp} is loosely defined as any feedforward \ac{ann} that contains at least three layers~\cite{xie2021segformer}.
It is composed of one input layer, one output layer, and at least one hidden layer:

\begin{itemize}
    \item Input layer: The layer where the input data is fed to the network, and the number of neurons on this layer is the same as the problem's dimensionality.
    \item Hidden layer: A function is applied to the previous layer's output, followed by an activation function. Many hidden layers can be stacked together.
    \item Output layer: The layer that produces the output values for the specific task the network was trained for. The number of neurons specified should be the same as the number of outputs.
\end{itemize}

For the network to produce accurate outputs, it must undergo a training process. The first step is the feedforward which is similar to an inference. After the data is fed to the network, the output is compared to the ground-truth values, often using a loss function. In our case, the error is calculated through the difference between the ground-truth and the output produced by the network.

\begin{figure}[t]
\centering
\includegraphics[width=0.9\textwidth]{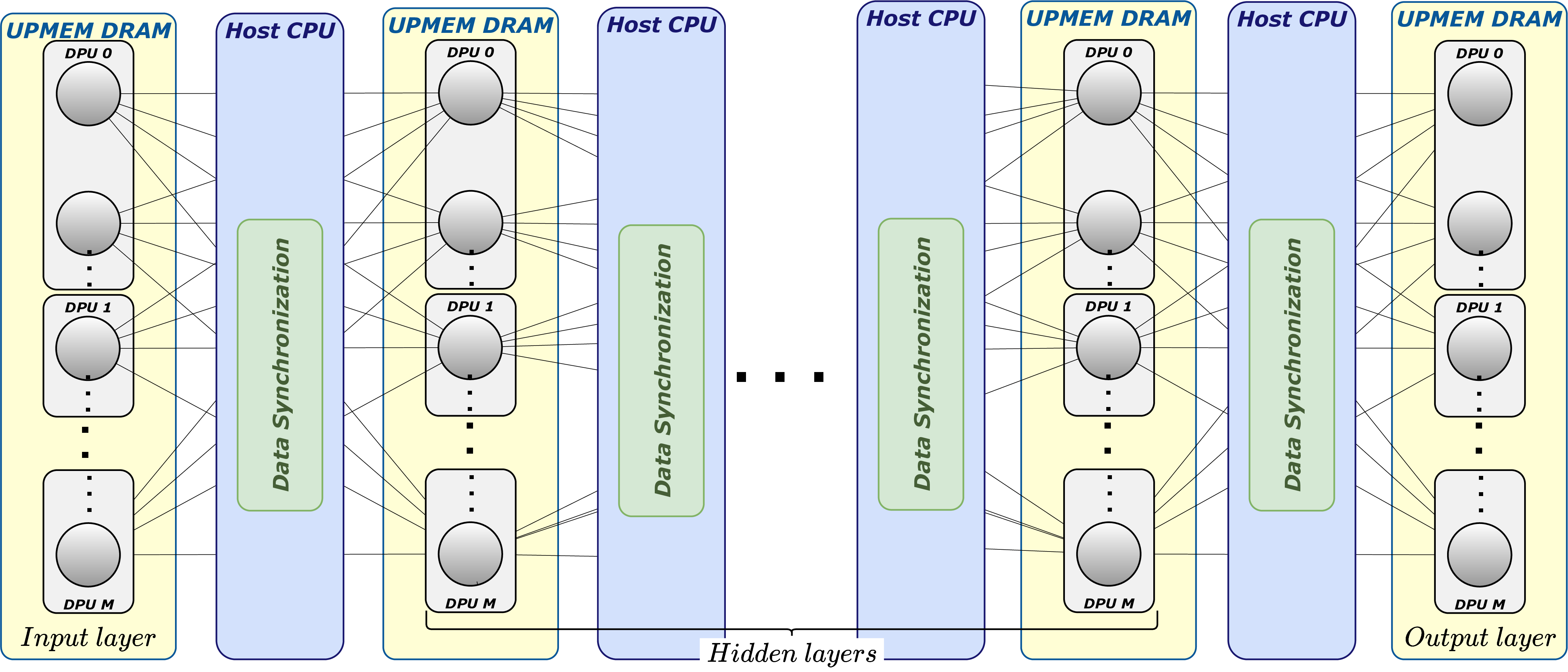}
\caption{Implementation of the UPMEM \ac{mlp}. Several \acp{dpu} processes multiple neurons. After executing all neurons in a layer, the data is synchronized by the \ac{cpu} and sent back to the \acp{dpu} for further processing.}
\label{fig:mlp_}
\end{figure}

The next step in training is backpropagation. Usually, an optimizer such as stochastic gradient descent is used~\cite{liu2020improved}, a learning rate is defined, and often other parameters for the optimizer are defined. The current case implements layers that calculate the derivative of the sigmoid function. The results are multiplied by a learning rate parameter when updating the weights.
Fig.~\ref{fig:mlp_} showcases how the \ac{mlp} is implemented across the \ac{pim} system.

\section{PIM-based MLP Approach}

We implemented training and inference in the UPMEM system.
Since the main focus of this work is inference, it is implemented using multiple \acp{dpu} and  multi-tasklets to measure the inference speed in different \ac{mlp} configurations.
Regarding training, a single \ac{dpu} was used with multi-tasklets to generate the weights to evaluate the model's accuracy, thus proving a correct implementation.
Furthermore, a \ac{gpu}-based implementation using \ac{cuda} of a smaller \ac{mlp} for inference on a low-power \ac{gpu} was used to compare this same network with a \ac{wram} implementation on \ac{pim}.

\subsection{Training}

In order to verify the model's accuracy, the Iris dataset was used~\cite{fisher1936use}. The Iris dataset consists of $150$ entries of iris flowers, each with four features (sepal length, sepal width, petal length, and petal width), and with each flower belonging to a different species (setosa, versicolor, and virginica). We conducted the training on $122$ entries and testing in the remaining $28$. The \ac{mlp} implementation aimed to classify between setosa and non-setosa samples.

The training was implemented in order to generate weights for the Iris dataset, confirming the correct function of the network, and it runs on a single \ac{dpu} with multi-tasklets.
For training, some additional kernels were defined.
The kernels for the feedforward pass and for the inference are the same.
For the backpropagation, several kernels were developed: 1) one kernel to calculate the sigmoid derivative, 2) one matrix subtraction kernel to calculate the error between the ground truth and the generated outputs, and 3) one element-wise matrix multiplication kernel to compute the element-wise product of two matrices, which is used to propagate the gradients backward through the network.

An input layer with $4$ neurons is used (since the Iris dataset has $4$ attributes). There is a single hidden layer with $8$ neurons and an output layer with a single neuron that specifies if the species of the iris is setosa or non-setosa.

\subsection{Inference}\label{sub_inf}

Regarding inference, two different \ac{dpu} kernels are required: 1) a kernel to implement parallel matrix multiplication using \ac{mram}, and 2) an activation function (sigmoid and ReLU) kernel that is applied to each block that constitutes the result of the multiplication, before retrieving the partial results back to the host. For this two approaches were taken, using either \ac{mram} or \ac{wram}.

\subsubsection{Matrix Multiplication}

In our implementation, we consider two scenarios where the multiplication of two large matrices that do not entirely fit into the \ac{wram} \ac{mram} of a single \ac{dpu}.
Therefore, the proposed solution employs a partitioning strategy, dividing the matrices into several blocks  that either fit in the \ac{mram} but not in the \ac{wram}, or that fit into the \ac{wram} to perform distributed \ac{gemm} using the UPMEM system.

For the matrix multiplication, we consider the first matrix to be in row-major order and the second one to be in column-major order. This optimizes host-device data transfers since now the data accesses of the first and second matrix are contiguous, and we use horizontal padding to ensure that the blocks are all of the same size to (which is a requirement for UPMEM's parallel transfers), and to be a multiple of $8$ bytes.
Also, since the data in each \ac{mram} bank is only accessible by a single \ac{dpu}, blocking is done at the granularity of entire rows, as data dependencies between different layers would require sending the data back to the host for synchronization, before sending the data again to diferent \acp{dpu}, as depicted in Fig.~\ref{fig:mlp}.

The UPMEM system's \ac{dma} engine requires the transfer size to be a multiple of $8$ bytes, which should also be considered when choosing the matrix dimensions and applying padding. For this reason odd numbers of rows should be avoided to mitigate the possibility of having chunks which are not multiple of $8$ bytes.

The fact that the second matrix is in column-major order allows the usage of padding with zeros by rows, which is faster than using column padding, since in the former case the memory addresses where the zeros are places are contiguous.
In addition, a larger portion of the bandwidth is utilized in the transfers since the \ac{dma} engine forces the use of contiguous blocks of transfers, allowing us to send entire rows in the second matrix at a time, instead of sending chunks of rows.
Also, since it is desirable to access the second matrix by columns for multiplication, the data would be properly aligned, and therefore the accesses would be faster.
For the second matrix (matrix $\textbf{B}$), padding is added to ensure that the blocks are of the same size and that the number of rows is even, to guarantee that the transfer size is a multiple of $8$ bytes.

\begin{figure*}[t]
\centering
\includegraphics[width=\textwidth]{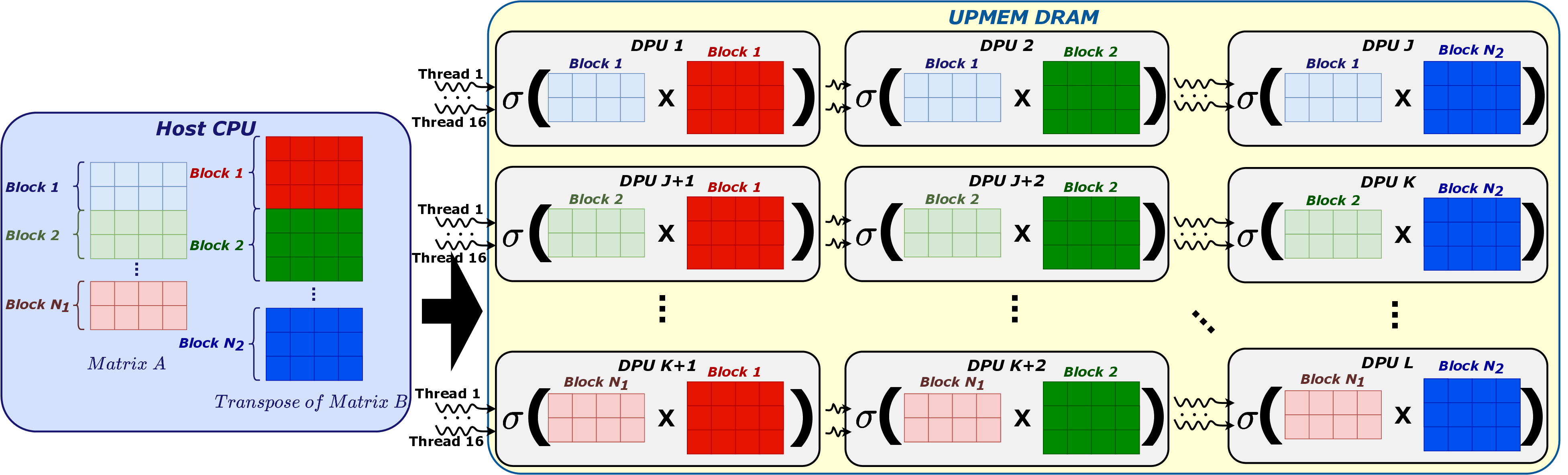}
\caption{Representation of the matrix multiplication in the UPMEM system. In this case, the total number of \acp{dpu} allocated is equal to $L$ and the processing is divided among the available. The $\textbf{B}$ is transposed in host \ac{cpu} to allow the transfer of contiguous block of data to the \acp{dpu}. Each block of matrix $\textbf{A}$ is replicated $N_2$ times in the \acp{dpu}, while the blocks of matrix $\textbf{B}$ are replicated $N_1$ times.}
\label{fig:parallelism_figure}
\end{figure*}

The allocation of the \acp{dpu} is performed once at the beginning of the program. When choosing the number of \acp{dpu} to allocate, we use one variable ($N_1$) for the number of blocks in the matrix $\textbf{A}$ and another variable ($N_2$) for the number of blocks in the matrix $\textbf{B}$.

Fig.~\ref{fig:parallelism_figure} showcases how the matrix multiplication is implemented.
First, each matrix is split into $N_1$ and $N_2$ blocks and sent to the \acp{dpu}.
In each \ac{dpu}, the blocks are multiplied and the results are sent back to the host.
The total number of \acp{dpu} allocated ($N$) should be the product of the number of blocks in each matrix:

\begin{equation} 
N_1\times N_2=N.
\end{equation}

The $N_1$ and $N_2$ variables should be positive integers within the range of the following expression:

\begin{equation} 
1\leq N_1, N_2\leq N.
\end{equation}

To provide a full matrix multiplication without partial results, the blocks of matrix $\textbf{A}$ must be replicated $N_2$ times in the allocated \acp{dpu} and matrix $\textbf{B}$ must be replicated $N_1$ times. This approach increases the total memory usage in the UPMEM system, which is highly desirable for a \ac{pim}-based application. The memory replication rate ($R$) should be as low as possible and can be modeled as follows ($\dim$ represents the size of the matrix): 

\begin{equation}
R(\%)=\left (  \frac{\dim(\textbf{A})\times N_2+\dim(\textbf{B})\times N_1}{\dim(\textbf{A})+\dim(\textbf{B})}\right )\times 100.
\end{equation}

Each \ac{dpu} launches $T$ threads (in these experiments $T=16$) with each processing $T_{rows}$ rows of the block, obeying to the following expression, where $C$ is the total number of rows of matrix~$\textbf{A}$: 

\begin{equation}
T_{rows}=\left \lceil \frac{\left ( \frac{C}{N_1} \right )}{T} \right \rceil. 
\end{equation}

\subsubsection{Activation Functions}

\begin{figure*}[!t]
\centering
\includegraphics[width=\textwidth]{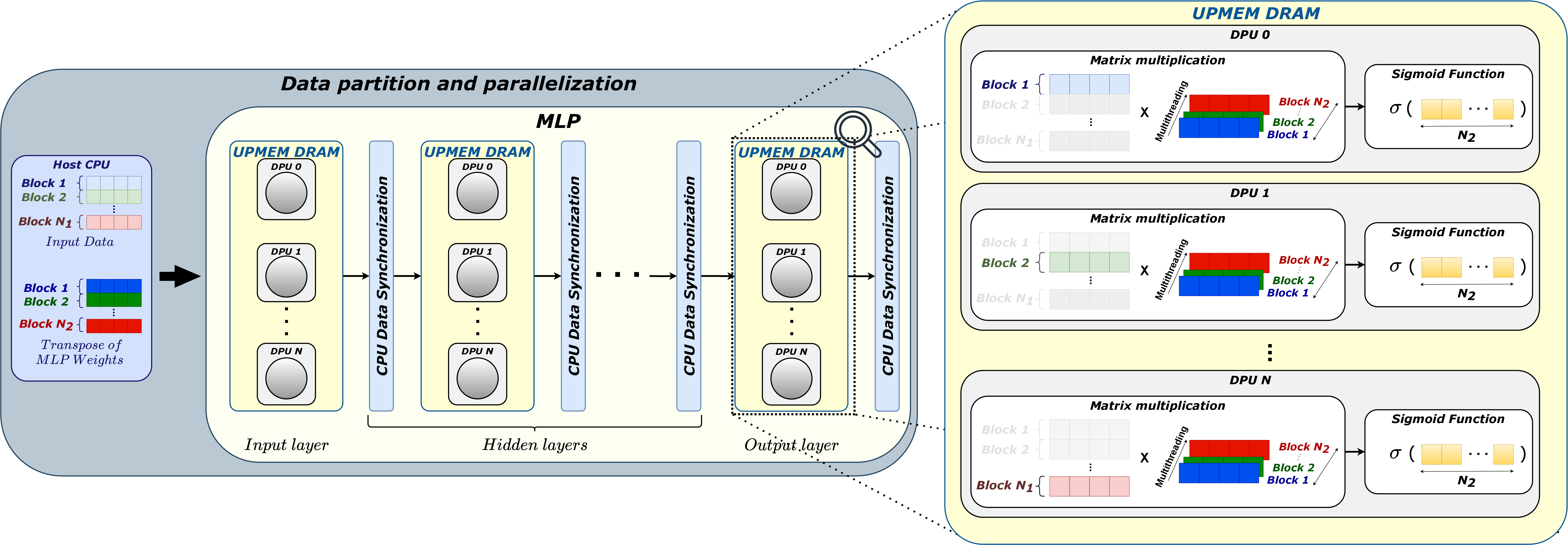}
\caption{Implementation of the UPMEM \ac{mlp}. On the left side of the figure, several \acp{dpu} process multiple neurons. After executing all neurons in a layer, the data is synchronized by the \ac{cpu} and sent back to the \acp{dpu} for further processing. The zoomed part on the right represents the matrix multiplication running in the UPMEM system. The total number of \acp{dpu} allocated is equal to $N$ and the processing is distributed among them. $\textbf{B}$ is transposed in the host \ac{cpu} to allow the transfer of contiguous blocks of data to the \acp{dpu}. Each block of matrix $\textbf{A}$ is replicated $N_2$ times in the \acp{dpu}, while the blocks of matrix $\textbf{B}$ are replicated $N_1$ times.}
\label{fig:mlp}
\end{figure*}

The sigmoid and ReLU are used as activation functions (represented as $\sigma$ in Fig.~\ref{fig:parallelism_figure} and Fig.~\ref{fig:mlp}). After sending the blocks of the matrices into the \acp{dpu} and performing the multiplication, the activation functions are applied directly to each element of each matrix block before sending the results back to the host.
The ReLU function is implemented using a comparison. The sigmoid uses an exponential function, and since  the UPMEM system does not support the standard C math.h library, the exponential is implemented according to~\cite{exp_macro}, which approximates the exponential from a double-precision float to a $32$-bit integer. Fig.~\ref{fig:mlp} maps the whole implementation in the full system.

\section{Experimental Results and Discussion}

 The \ac{pim}-based experiments were conducted in the UPMEM servers with $20$ \ac{pim} \acp{dimm}, totaling $2560$ \acp{dpu} (in UPMEM's servers available for conducting our experiments, some \acp{dpu} were not accessible), each running at $350$ MHz, and $160$ GB of \ac{pim} memory. The available \ac{cpu} is the Intel Xeon Silver $4215$ \ac{cpu} @ $2.50$GHz~\cite{upmem_system}. 
 Regarding the low-power \ac{gpu} experiments, they were conducted on the Jetson AGX Xavier~\cite{AGX_Deploy} (Volta architecture), with $512$ \ac{cuda} cores, and $48$KB of programmable L1/shared-memory per block. It operates at up to $1377$MHZ with a power budget of $30$ W, providing up to $32$ \ac{tops}.

To compare against the low-power \ac{gpu}, we implemented the inference kernels described in subsection~\ref{sub_inf}, using \ac{cuda} and the cuBLAS library, Since smaller matrix dimension were chosen, this approach allowed the blocks to be transferred directly without additional padding, which serves as a mean to compare with the UPMEM-based approaches using \ac{mram} and \ac{wram}.

\subsection{Evaluating Model Accuracy}

To validate the classification accuracy of a model generated by the UPMEM system, we first conducted a training run using a single \ac{dpu} with multithreading to generate a model with trained weights.
The Iris dataset is split into a training and test dataset.
The test dataset consists of $8$ random samples of iris-setosa, $10$ of iris-versicolor, and $10$ of iris-virginica.
We then set iris-setosa to $0$ and iris-versicolor and iris-virginica to $1$ as the goal was to classify between iris-setosa and not iris-setosa.
To conduct training, the remaining $122$ inputs were used, setting the batch size to $122$, using $0.1$ as the learning rate and conducting training for $500$ epochs.
During all the conducted experiments, $16$ tasklets were used, as using a power of $2$ would avoid misalignment issues that could have occurred.
As stated in~\cite{prim_other}, the arithmetic throughput of a \ac{dpu} is saturated at $11$ or more tasklets, and thus there is no improvement when using more than $11$ threads.
Using the weights generated during training, the model managed to correctly label all $28$ samples in the testing dataset, thus achieving $100\%$ accuracy in the test dataset.

\begin{table*}[t]
       \centering
            \caption{Sizes of the inputs, layers, and outputs of the networks used to assess inference speed.}
        \label{network_dim}
            \resizebox{0.8\linewidth}{!}{%
            \begin{tabular}{cccccc}
            \toprule
            \textbf{Network} & \textbf{Input} & \textbf{Input layer} & \textbf{Hidden layer 1} & \textbf{Hidden layer 2} & \textbf{Output} \\ \toprule
        $Net_1$ (LeNet5-based) & 9984 & 512 & 128 & 64 & 1 \\ \hline
        $Net_2$ (VGG-based) & 16384 & 16384 & 4096 & 4096 & 1 \\ \hline
        \multirow{5}{*}{$Net_3$ (LeNet5-based)}  & 2556 & \multirow{5}{*}{112} & \multirow{5}{*}{96} & \multirow{5}{*}{64} & \multirow{5}{*}{1} \\ \cmidrule(l){2-2}
         & 5112  &  &  &  &  \\ \cmidrule(l){2-2}
         & 7668  &  &  &  &  \\ \cmidrule{2-2}
         & 10224 &  &  &  &  \\ \cmidrule{2-2}
         & 15336 &  &  &  &  \\ \hline
        \multirow{3}{*}{ $Net_4$ (VGG-based)}  & 2556 & \multirow{3}{*}{176} & \multirow{3}{*}{64} & \multirow{3}{*}{64} & \multirow{3}{*}{1} \\ \cmidrule{2-2}
         &   5112 &  &  &  &  \\ \cmidrule{2-2}
         &   7668 &  &  &  &  \\ \hline
            \end{tabular}
            }
            
\end{table*}

\subsection{Evaluating Inference Performance on UPMEM vs CPU}

To assess the performance improvement of the UPMEM system relative to single-threaded \ac{cpu} execution, the results obtained were compared by varying the number of \acp{dpu} for different inputs and configurations.
The data types used are 32-bit floating point (FP32) and 32-bit integer (INT32).
Since the UPMEM system does not have hardware support for FP32 operations, nor for INT32 multiplications, the system relies on emulation that uses integer-specific hardware to support them.
Although we used the INT32 format, some of the required operations must occur using floating-point precision (e.g., exponentiation and the \texttt{ceil} function).
The implementation was tested in real-case scenarios.
For this reason, we set the number of layers, neurons, inputs, and outputs to match the number of fully connected layers in the LeNet5~\cite{lenet5_paper} ($Net_1$) and  VGG~\cite{vgg_paper} ($Net_2$) networks.
To avoid 8-byte alignment issues, we chose layer sizes close to those of LeNet$5$ and VGG but using either powers of $2$ or combinations of powers of 2.
The values for the inputs and weights were randomly generated as we only intend to assess inference speed.
Regarding the input size, we chose values close to those of the MNIST testing dataset and the Cifar-$10$ testing dataset (both have $10000$ images).

Additionally, to simulate the activations in $Net_2$ (VGG-based), a ReLU function was implemented. The number of neurons for the input layers reflects the size of the last layer before the fully connected layers after flattening.
Table \ref{network_dim} displays the values used for each network.
The size of the testing dataset for MNIST is $10000$, and for ImageNet is $50000$, and we chose $9984$ and $16384$, respectively, for the current tests and scenarios.
Before the fully connected layers, a flatten operation must be applied to the output tensor from the last convolutional layer, therefore for $Net_1$, the number of neurons for the input layer is $5\times5\times16 = 400$, so $512$ was chosen, and for $Net_2$, the number of neurons is $7\times7\times512 = 25088$ and $16384$ was selected.
The hidden layer sizes for $Net_1$ are $120$ and $84$, so $128$ and $64$ were chosen. For $Net_2$, the sizes of both hidden layers are 4096, and since it is a power of $2$, we kept the size. For the output layers, even though the MNIST dataset has $10$ classes and ImageNet has $1000$ classes, we chose $1$ for the problem to remain a single-class problem, and the sigmoid function could be used instead of a softmax.

\begin{figure*}[!t]
\centering
\includegraphics[width=0.8\textwidth]{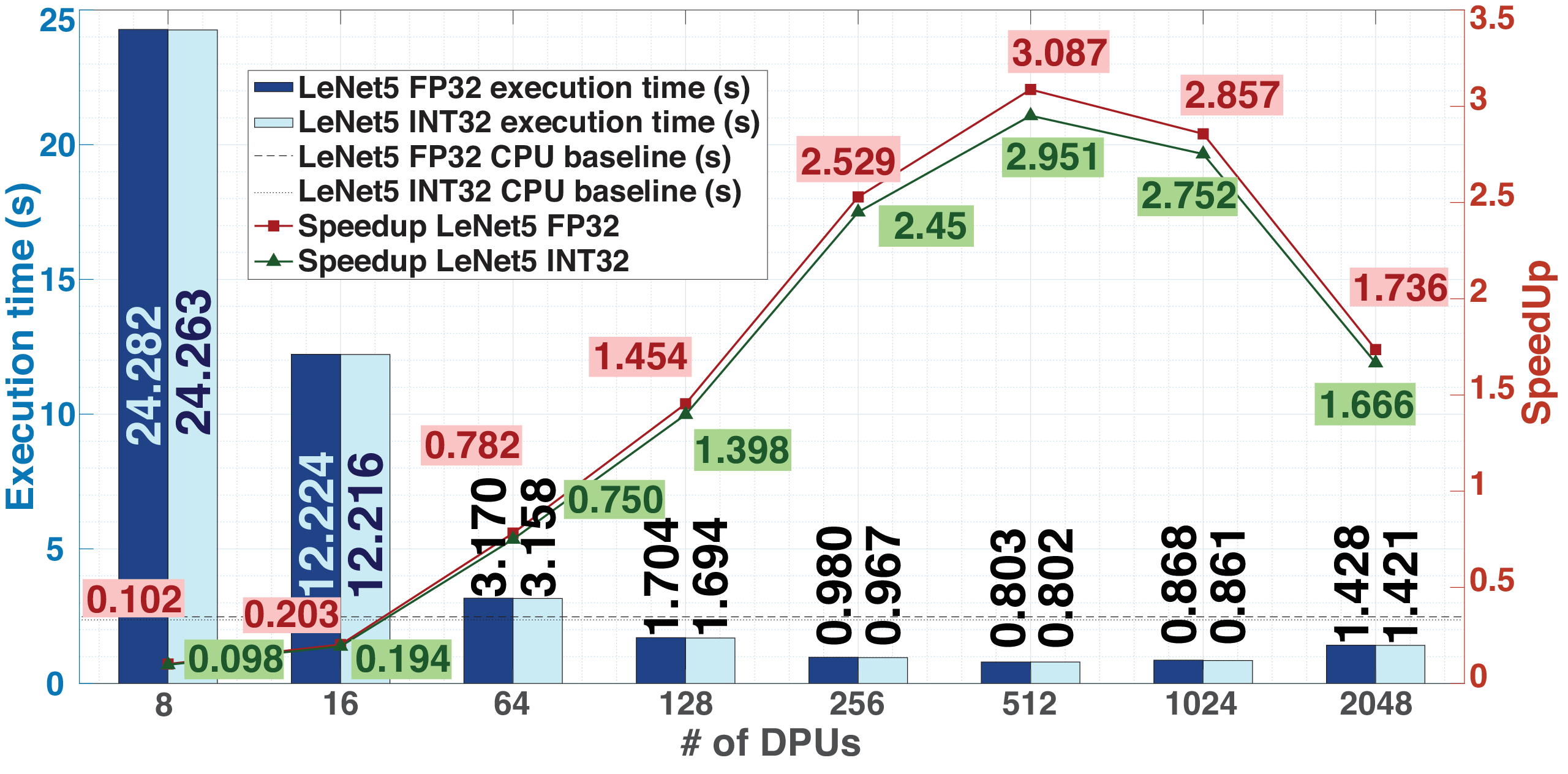}
\caption{Inference time and speedup compared to \ac{cpu} of $Net_1$ for FP$32$ and INT$32$ representations.}
\label{fig:times_lenet_arrow}
\end{figure*}

\begin{figure*}[!t]
\centering
\includegraphics[width=0.8\textwidth]{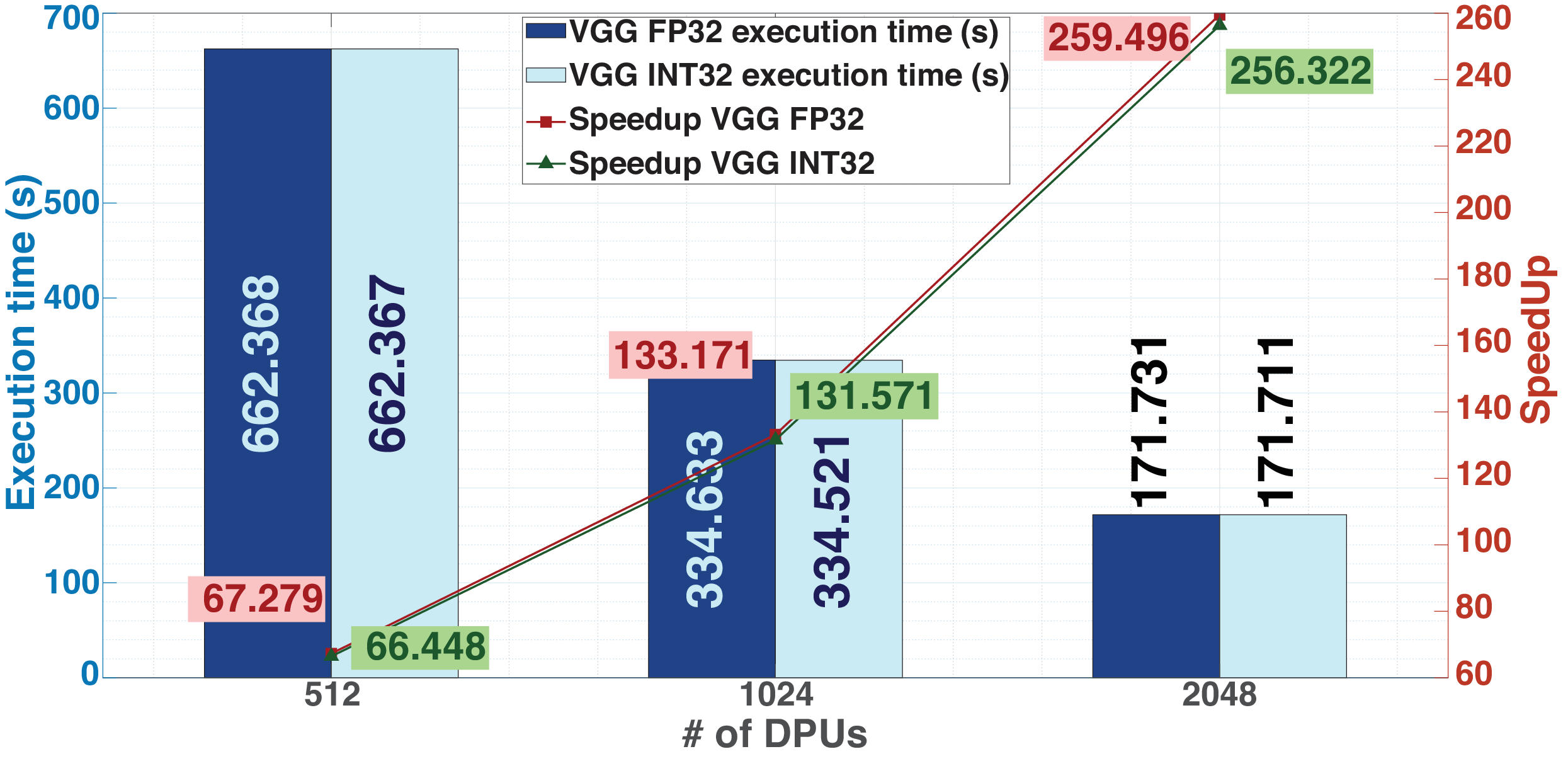}
\caption{Inference time and speedup compared to \ac{cpu} of $Net_2$ for FP$32$ and INT$32$ representations.}
\label{fig:times_vgg_arrow}
\end{figure*}

Fig.~\ref{fig:times_lenet_arrow} showcases the comparison of the network on different numbers of \acp{dpu} compared to the \ac{cpu} sequential baseline both for FP32 and INT32. We can see that for the $Net_1$, the best time both for the FP32 version and INT32 version is achieved using 512 \acp{dpu}. This is because for the amount of data available, allocating more \acp{dpu} incurs overheads that do not reflect a speedup since the allocation operation is expensive, and using more \acp{dpu} may incurs additional padding that is also costly.
We can observe that the speedup is higher for the FP32 version.
This is because even though the INT32 time is lower in all cases, the speedup is calculated against the sequential times that were obtained also using INT32 and FP32.
These times are different from each other, with the INT32 time being slightly lower, as can be seen in Fig.~\ref{fig:times_lenet_arrow}. 

The sequential times obtained for the FP32 and INT32 were \SI{2479}{\milli\second} and \SI{2368}{\milli\second}, respectively. The best time obtained for the FP32 was \SI{803}{\milli\second}, corresponding to a $3\times$ speedup, and for the INT32 version was \SI{802}{\milli\second} which corresponds to a $2.9\times$ speedup. To obtain the results, 6 repetitions after 5 warm-ups were performed, and the average was calculated. 

Fig.~\ref{fig:times_vgg_arrow} depicts the results obtained for $Net_2$. We can observe that the best inference speed is obtained using 2048 \acp{dpu}, which is the maximum number possible.
The times for the sequential version at FP32 and INT32 were \SI{44563}{\second} ($\sim$\SI{743}{\minute}) and \SI{44013}{\second} ($\sim$\SI{734}{\minute}), respectively.
The best times obtained for FP32 and INT32 were, respectively, \SI{171.73}{\second} ($259\times$ speedup) and \SI{171.71}{\second} ($256\times$ speedup).

\subsection{Evaluating Inference Performance on UPMEM vs a Low-power GPU}

To provide a comparison with low-power \acp{gpu} two smaller networks were created. The first network, denoted as $Net_3$, consists of an input layer with $112$ neurons, followed by two hidden layers with $96$ and $64$ neurons, respectively, and an output layer with a single neurons. The second network, $Net_4$, has an input layer with $176$ neurons, followed by two hidden layers with $64$ neurons each, and an output layer with a single neuron. 

In order to assess the maximum inference performance of the \ac{mlp} on \ac{wram}, all available \acp{dpu} were utilized, and matrix $B$ was entirely replicated across each \ac{dpu}, thus avoiding partitioning of matrix $B$ (Fig.~\ref{fig:parallelism_figure}).


However, host-to-\ac{wram} transfer speed is significantly slower than host-to-\ac{mram} transfer due to the system’s memory architecture. \ac{mram} is the UPMEM system’s primary memory, connected to the host via the \ac{dram} bus, enabling high-speed data transfers~\cite{lee_2024_A_Memory_Management}. In contrast, \ac{wram} is not directly accessible by the host~\cite{upmem_system}. When transferring data to \ac{wram}, the host must first write to \ac{mram}, after which \acp{dpu} must copy the data into \ac{wram}, introducing additional overhead~\cite{In_Memory_Acceleration}. \ac{wram} is designed for fast on-chip processing, rather than bulk transfers, and its bandwidth is constrained by internal \ac{dpu} memory limitations, whereas \ac{mram} benefits from high-bandwidth \ac{dram} interfaces.

The selected batch sizes were the largest that could fit within each \ac{dpu}'s \ac{wram} without requiring frequent transfers between \ac{mram} and \ac{wram}.

\begin{figure*}[!t]
\centering
\includegraphics[width=0.8\textwidth]{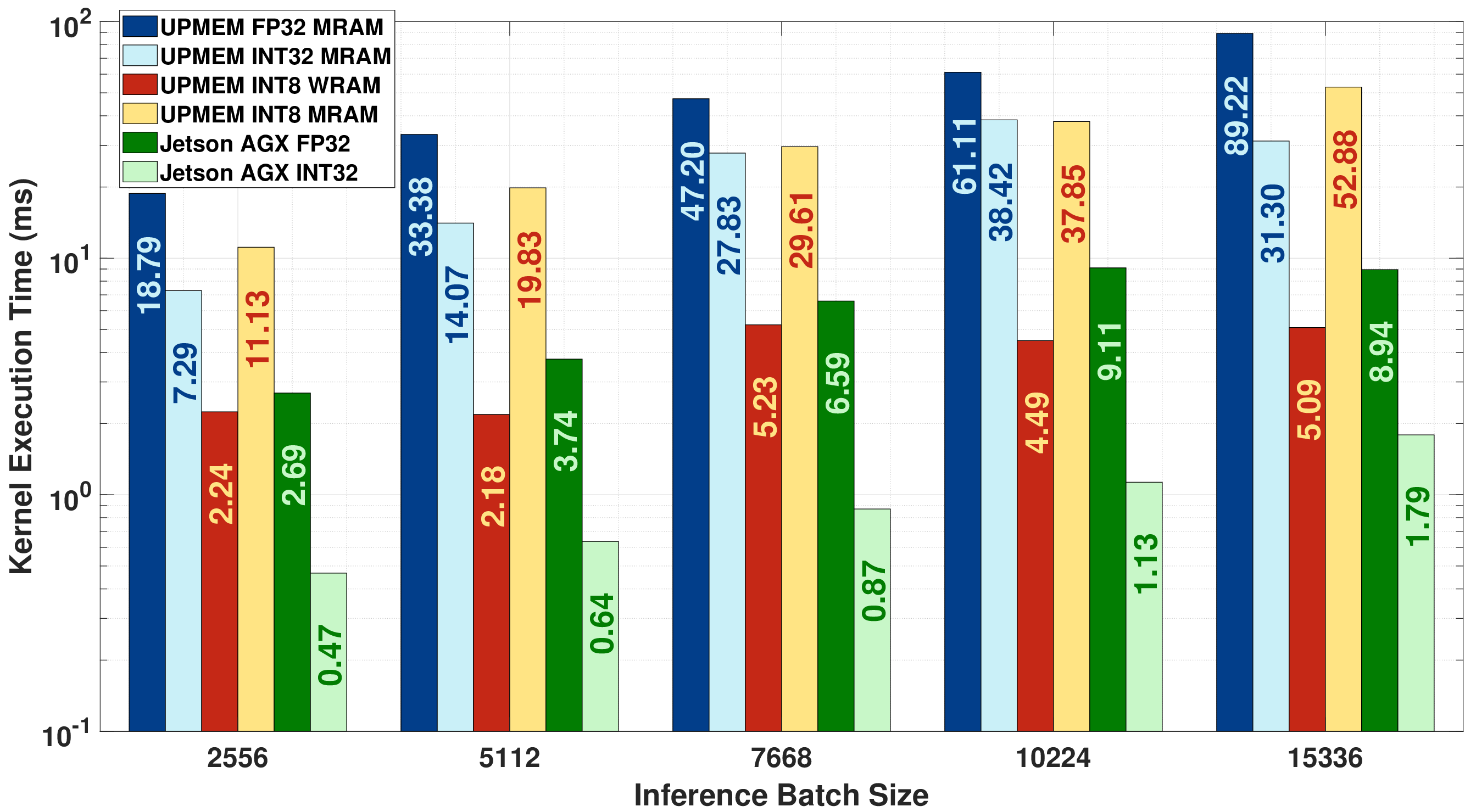}
\caption{Comparison of kernel execution time across different inference batch sizes used on $Net_3$ in the UPMEM system and AGX Xavier GPU.}
\label{fig:times_kenel_net5}
\end{figure*}

\begin{figure*}[!t]
\centering
\includegraphics[width=0.8\textwidth]{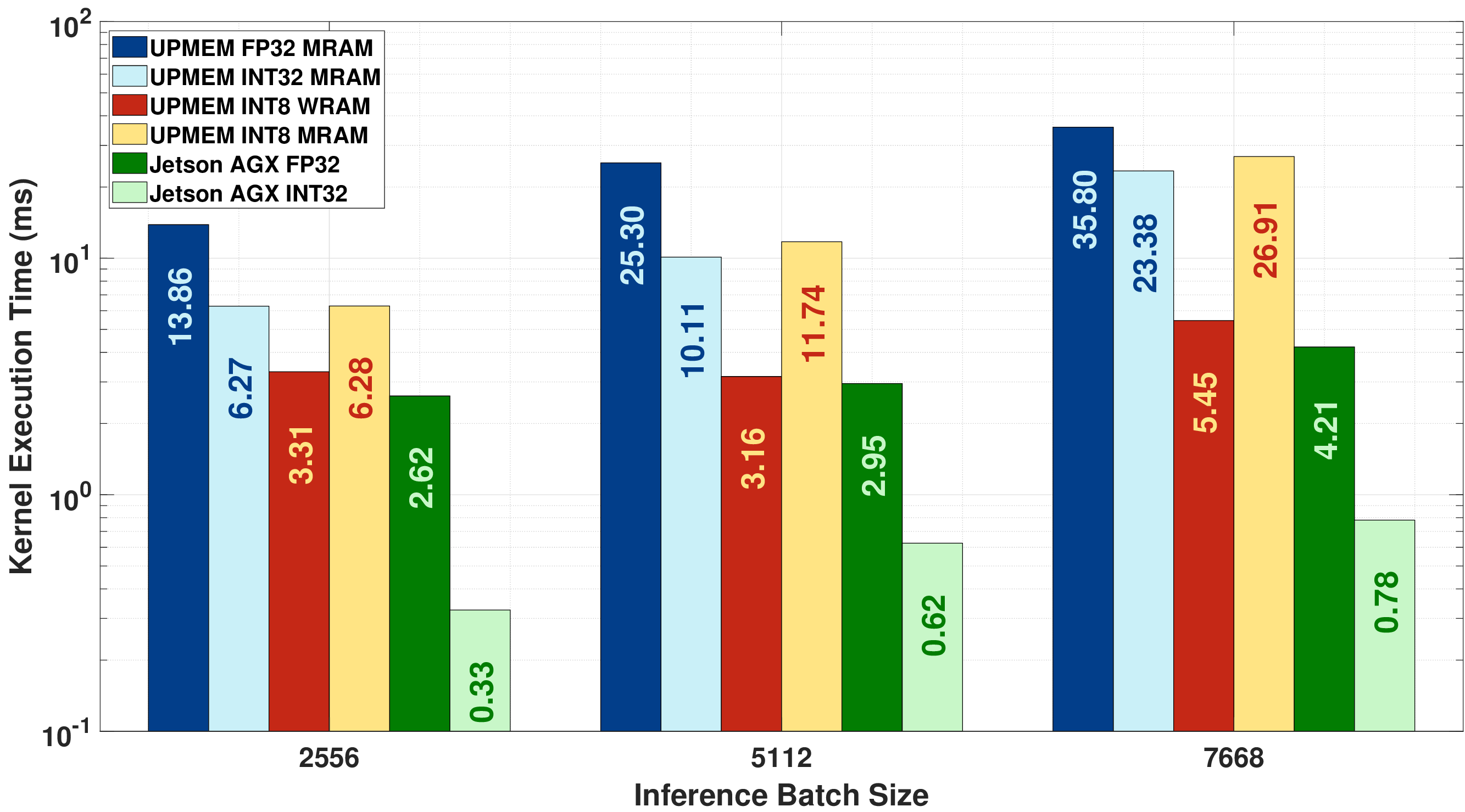}
\caption{Comparison of kernel execution time across different inference batch sizes used on $Net_4$ in the UPMEM system and AGX Xavier GPU.}
\label{fig:times_kenel_vgg}
\end{figure*}

The kernel execution time (Fig.~ \ref{fig:times_kenel_net5} and Fig~~\ref{fig:times_kenel_vgg}) indicates that utilizing WRAM for processing in UPMEM yields shorter execution times than MRAM, as anticipated due to diminished memory access latency. This effect is especially prominent in 8-bit MLP, where WRAM markedly enhances kernel execution speed. The observed disparity between FP32 and INT32 operations indicates that, although both are emulated, INT32 entails reduced overhead, resulting in enhanced performance.

\subsection{Assessing the Impact of Data Transfers}

\begin{figure}[t]
    \centering
    \begin{subfigure}[t]{0.49\linewidth}
        \centering
        \includegraphics[width=\linewidth]{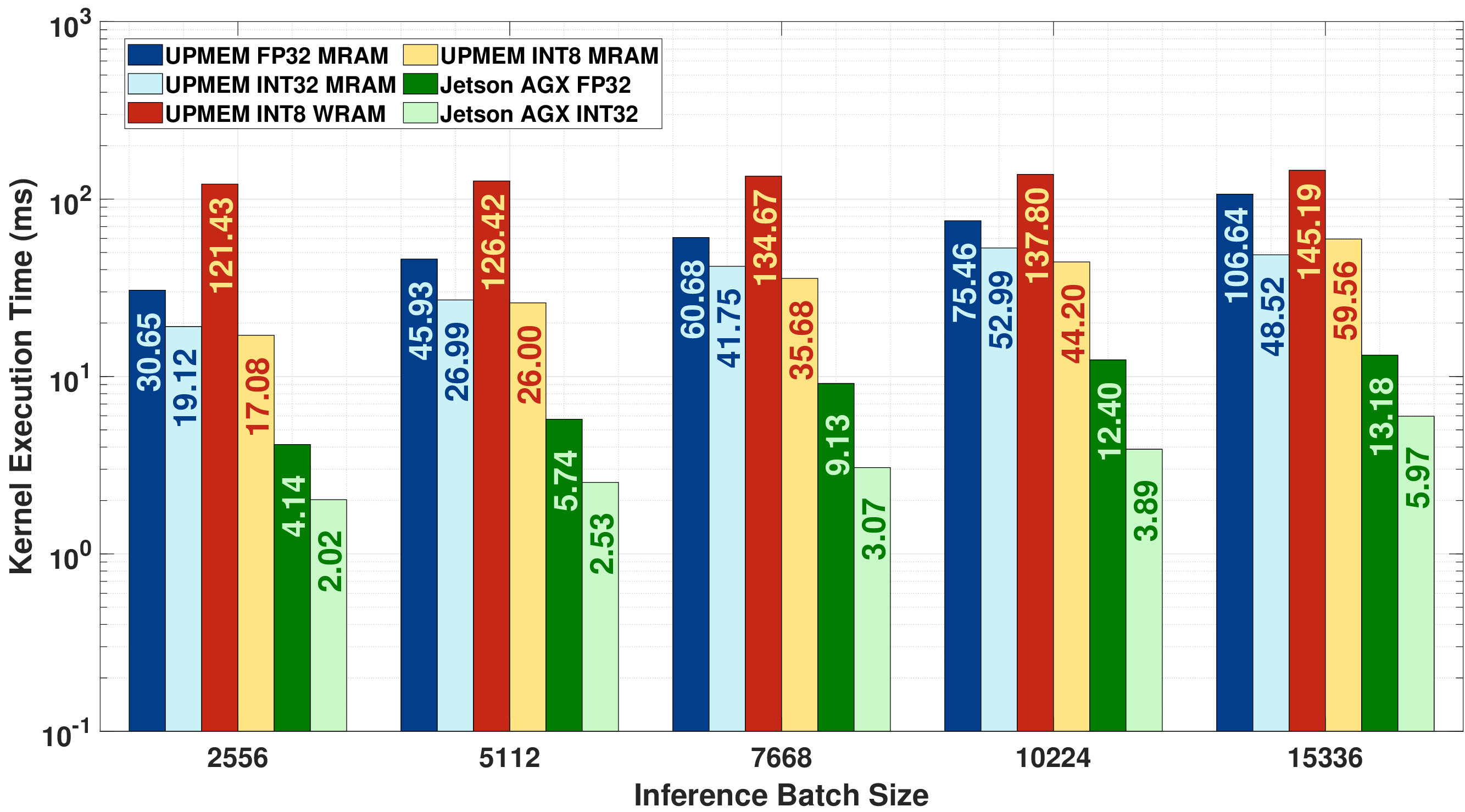} 
        \caption{$Net_3$ total execution time(including data transfers).}
        \label{fig:times_Total_net5}
    \end{subfigure}
    \hfill 
    \begin{subfigure}[t]{0.49\linewidth}
        \centering
        \includegraphics[width=\linewidth]{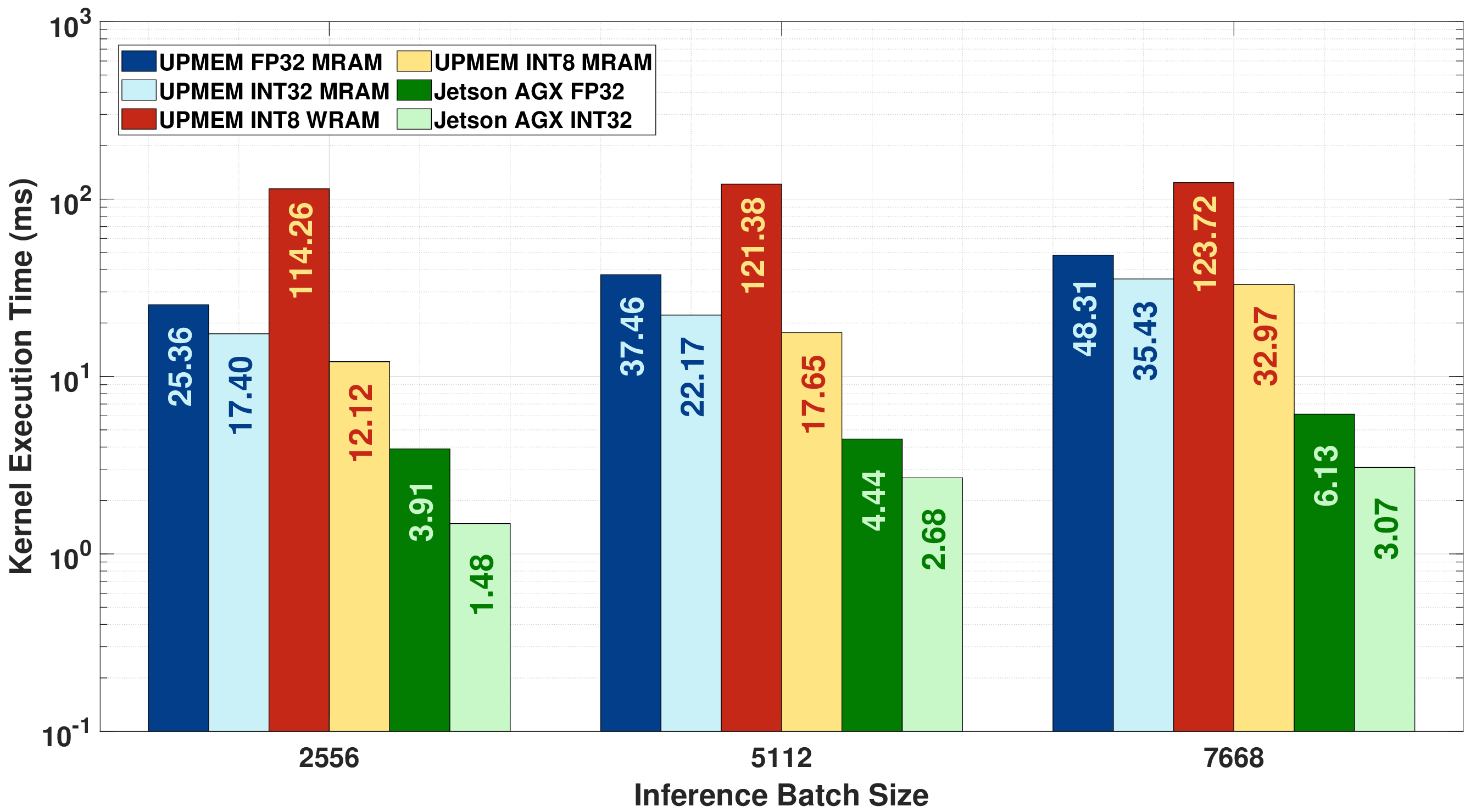} 
        \caption{$Net_4$ total execution time (including data transfers).}
        \label{fig:times_Total_vgg}
    \end{subfigure}
    
    \caption{Evaluation of total execution time (kernel execution time and data transfers) for different inference batch sizes in $Net_3$ and $Net_4$, implemented on the UPMEM system and AGX Xavier GPU. The results highlight the influence of memory hierarchy and data movement on execution performance.}
    \label{fig:times_kerne_vs_total}
\end{figure}

The negative impact of data transfers on execution time in UPMEM systems is clearly illustrated in Fig.~\ref{fig:times_Total_net5} and Fig.~\ref{fig:times_Total_vgg}, where the overall execution time, with data transfers, of the UPMEM system and AGX Xavier is compared across various batch sizes and two separate neural networks.  Although \ac{pim} systems aim to reduce data movement by executing calculations directly in memory, unavoidable data transfers in this case nevertheless incur substantial overhead, especially when relating with host memory or synchronizing across different memory areas. The charts indicates that AGX Xavier results (FP32 and INT32) exhibit markedly reduced execution times relative to \ac{pim}-based implementations, frequently by an order of magnitude (roughly 10× faster).  This underscores the significant consequences of data transfer penalties in contemporary \ac{pim} architectures.


A notable observation is that the UPMEM INT8 \ac{wram} consistently demonstrates the longest execution times across all batch sizes compared to INT8 \ac{mram}, indicating that, despite employing \ac{pim}, the memory transfer overheads in \ac{wram} remain considerable.  This indicates that \ac{mram} is a more effective option for execution when low data re-utilization exist, whereas \ac{wram} should be circumvented, i.e. when the problem is small enough. The scalability trend 
based on the derivative of total execution time with respect to inference batch size indicates that AGX Xavier manages larger batch sizes more effectively, with execution time increasing more gradually in comparison to UPMEM-based implementations. Conversely, UPMEM execution time grows noticeably with increasing batch size, indicating that data transfer and memory access latencies become progressively more problematic at bigger sizes.

In summary, whereas \ac{pim}-based execution is anticipated to diminish data transmission cost by executing computations directly in memory, the present generation of UPMEM systems continues to experience considerable penalties related to data movement, especially in WRAM-based configurations.  Future \ac{pim} architectures must enhance memory access strategies, reduce unneeded host interactions, and augment data localization to alleviate these inefficiencies.  With the emergence of next-generation \ac{pim} solutions, we expect these overheads to decrease, allowing \ac{pim} to compete more effectively with conventional GPU-based execution models.

\section{Related Work}


In a processor-centric architecture, the processing unit (\ac{cpu}) is decoupled from the memory. In order to alleviate the data movement bottleneck, the data is forced to circulate over a memory hierarchy (cache), where prefetchers are used. This mechanism increases the system's complexity and some workloads cannot exploit temporal or spatial locality to increase performance~\cite{big_data_analitics, google_consuptions, google_warehouse, pim_workload_driven_prespective}.

\ac{pim} is an emerging paradigm that proposes to process data 
 near or where it resides, i.e., in the memory system~\cite{modern_primer, processing_data_where_it_makes_sense}. This is accomplished by adapting \ac{dram}'s control electronics technology that manipulates data in the memory cell (\ac{pum}) or by placing computing units near the memory system (\ac{pnm}). 

The most relevant technologies used for the processing elements in \ac{pim} available in the literature can be found in Table~\ref{tab:pim}, where some of the references refer to simulators and frameworks that do not directly offer standalone solutions in \ac{pim} hardware, but offer a way to benchmark or evaluate \ac{pim} systems. Even though UPMEM offers the only-commercially available \ac{pim} solution, there have been other general-purpose \ac{pim} solutions that are not commercially available, for example, Samsung released  AXDIMM~\cite{samsung_axdimm}, which is an \ac{fpga}-based DDR4 compatible platform that includes both general-purpose processing units as well as dedicated processing units for deep learning, where the host can directly access the \ac{dram} banks in the platform or issue instructions to execute in-memory computations using the \ac{dram} ranks inside the device. Samsung's HBM-PIM~\cite{samsung_hbm_pim} is a platform that is composed of 3D-stacked \ac{dram} banks (although other \ac{dram} standards also work), and \ac{pim} units that execute \ac{fp16} operations such as multiplication, addition, move, \ac{mad}, and \ac{mac}, where the host \ac{cpu} issues commands to the device. AiM~\cite{AiM_paper} is based on GDDR6 and is composed of 16 \ac{pim} processing units, one for each bank, which executes operations such as \ac{mac}, activation functions, element-wise multiplication, and other operations for deep learning specific applications. HB-PNM~\cite{HB-PNM_paper} is composed of \ac{dram} blocks that serve as memory for two computation engines.

\begin{table*}[t]
    \centering
            \caption{\ac{pim}-based works available in the literature}
            \label{tab:pim}
            \resizebox{\linewidth}{!}{ 
            \begin{tabular}{|l|l|}
            \hline
            \multicolumn{1}{|c|}{\textbf{Processing-near-Memory}} & \multicolumn{1}{c|}{\textbf{Processing-using-Memory}} \\ \hline
            \begin{tabular}[c]{@{}l@{}}General-purpose cores~\cite{syncron, pim_graph_processing, google_consuptions, conda, napel_essemble_learning, damov, heterogeneous_approach_acceleration}\\ Application-specific accelerators~\cite{pim_time_series, pim_genome_sequence,grim_dna_mapping_pim,jafar_database_pim, sparse_matrix_matrix_pim, data_reorg_pim, pointer_pim, stencil_pim}\\  Simple functional units (such as \ac{hmc})~\cite{ahn2015pim, graphpim, cairo}\\ \Acp{gpu}~\cite{tom,pim_gpu_standarize, scheduling_techniques_pim_gpu, top-pim, pnm_in_gpu}\\  Reconfigurable logic~\cite{chameleon,hrl,3d_stacked_fpga_pim}\end{tabular} & \begin{tabular}[c]{@{}l@{}}  \Ac{sram}~\cite{in_cache_computing, in_cache_nns, in_cache_paralelization, sram_embebed_computations,neurosim, bit_precision_nn_reconfigurable_architecture,zpim_accelerator_architecture}\\  \Ac{dram}~\cite{ambit2, rowclone, dram_rng, compute_dram, drisa_dram, simd_dram, rng_row_activating}\\  \Ac{nvm}~\cite{pinatubo, pima, cpim_hardware_accelerator_comparator, aligns, levy2014logic, kvatinsky2014magic, isaac_cnn_accelerator, kvatinsky2011memristor, kvatinsky2013memristor, gaillardon2016programmable, bhattacharjee2017revamp, hamdioui2015memristor, xie2015fast, yu2018memristive, prime_reram_accelerator, zheng2016rram, xi2020memory, puma, panther, emulater_for_ML, bruel2017generalize, huang2021mixed, floatpim_accelerator, parapim, neurosim} \end{tabular} \\ \hline
            \end{tabular}   
        }
\end{table*}

\subsection{Neural Networks using PIM}



In ISAAC~\cite{isaac_cnn_accelerator}, the authors present a \ac{pum} architecture based on memristor crossbar arrays for \ac{cnn} inference that exploits the characteristics of the memory to implement analog operations.
In FloatPIM~\cite{floatpim_accelerator}, the authors propose a \ac{pum} architecture that implements digital operations (thus eliminating \ac{adc} and \ac{dac} overheads) using memristors, and it supports inference and training of \acp{cnn} in floating-point precision.
ParaPIM~\cite{parapim}, uses \ac{sotmram} to enable inference of \acp{bnn}, where first the input feature maps and the weights are binarized and then mapped to the memory subarrays.
\Ac{sotmram} was used in CMP-PIM~\cite{cpim_hardware_accelerator_comparator} to implement a novel \ac{cnn}, in which the multiplications in the convolutions were replaced with comparisons and additions to achieve faster inference.
The authors of~\cite{heterogeneous_approach_acceleration} proposed a heterogeneous \ac{pim} framework for \ac{nn} training and inference, with both fixed function logic and programmable \ac{pim} logic, implemented in 3D-stacked memory. 
In~\cite{bit_precision_nn_reconfigurable_architecture}, a reconfigurable $1$ to $16$ bit \ac{pim} computing macro is proposed based on bit cells which are composed of a \ac{sram} cell for binary weight storing, an XNOR gate for bit-wise multiplication, and an adder for bit-wise addition.
The authors in~\cite{ferreira2022pluto}, propose pLUTo, a \ac{pum} technology based on look-up table operations and shows improvements for LeNet$5$ in terms of throughput and energy performance compared to main baselines.
\Ac{reram} technology was used in PRIME~\cite{prime_reram_accelerator}, to allow \ac{nn} acceleration using a developed software/hardware interface.
Z-PIM\cite{zpim_accelerator_architecture}, offers a fully variable weight bit precision based on \ac{sram} technology. Z-PIM is composed of a core, an input load unit, a weight load unit, and an output buffer that implements convolution and accumulation in memory.
The authors of~\cite{pnm_in_gpu} proposed NDPX, a novel architecture for accelerating memory-bound operations in deep neural network training, achieving a 51\% speedup for training VGG-16 network, by integrating \ac{pnm} logic with memory expanders and \ac{gpu} offloading, which allows for the execution of memory-bound operations by \ac{pnm} units.
In~\cite{hrl}, the authors proposed \ac{hrl}, which uses \ac{pnm} to combine both the power efficiency of an \ac{fpga}, and the area efficiency of an \ac{cgra}, which improved the performance of workloads such as graph processing, and \acp{nn}.
Das et al.~\cite{das2022implementation} explore the implementation and evaluation of \acp{dnn} on commercially available \ac{pim} hardware, specifically on the UPMEM architecture. They address key challenges of the UPMEM platform by adapting \ac{cnn} architectures via quantization, selective task delegation, and data padding techniques. The authors demonstrate the viability of running \ac{cnn} algorithms, achieving substantial performance gains over traditional \ac{cpu}-based implementations. This work highlights the potential and challenges of utilizing \ac{pim} hardware for accelerating machine learning workloads.

\subsection{PIM Workloads using UPMEM's System}

This subsection presents a brief overview of some papers, that we deemed as relevant that already use UPMEM's system.
In~\cite{prim, prim_other}, the authors use micro-benchmarks to test UPMEM's architecture limits, such as compute throughput and memory bandwidth, and benchmark 16 workloads such as vector addition, \ac{mvm}, sparse \ac{mvm}, select, time series analysis, multi-layer perceptron, matrix transposition, and others. We want to point out that the multi-layer perceptron does not perform inference or training of any size, and the batch size is always set to 1. This is because the goal was mainly to benchmark the multi-layer perceptron without having a functional \ac{nn}.
In~\cite{5_benchmark_upmem}, the authors test the computing capability and the memory bandwidth scaling by choosing five well-known workloads in which the computational resources increase with data size and have low data dependencies. These workloads were Snappy~\cite{snappy}, a hyperdimensional computing application, an advanced encryption standard application, JSON filtering, and Grep~\cite{grep}.
In~\cite{ml_pim, ml_pim_abstarct}, the authors made a \ac{pim} implementation of machine learning algorithms such as K-means clustering, linear regression, Logistic regression, and decision trees, analyzing them in terms of accuracy, performance and scaling against implementations in \ac{cpu} and \ac{gpu}.
In~\cite{sparse_matrix_pim}, a C library for Sparse matrix multiplication is created, where various compressed formats, load balancing schemes, and synchronization approaches are available. The implementations were also compared against \ac{cpu} and \ac{gpu}.
The authors of~\cite{pim_tree} created a skew-resistant index that dynamically decides either to push queries to the \ac{pim} node or to pull the node's keys back to the \ac{cpu} based on the workload skew.
In regards to Genomics and Bioinformatics, in~\cite{blast_upmem} a \ac{pim} implementation of BLAST~\cite{blast}, which is an algorithm to perform \ac{dna} or genomic protein banks scanning. In~\cite{variant_calling} \ac{pim} is used to accelerate the variant-calling genomic process, which is a genomic analysis using a novel parallelization technique suited for UPMEM's \ac{pim}. The authors of~\cite{genomic_framework} proposed a framework for performing genomic sequence alignment.



\section{Conclusions}
The need to explore alternative locations to place computing units assumes greater importance. Memory emerges as the natural candidate for core placement. However, \ac{pim} is not meant to be a substitute for existing accelerators or architectures. \Ac{pim} is supposed to be complementary to them. 
Using the UPMEM's system poses numerous challenges, such as more support for complex \ac{nn} implementations, convolution operation, and cache coherence mechanisms. More efficient multipliers that are not based on emulation are also needed.  
The number of \acp{dpu} used must be carefully chosen as allocating an excessive number might incur unnecessary allocation overheads that distribute the data over more \acp{dpu}. At the same time, there was enough space available at the \ac{mram} banks, and it could also cause excessive padding in order for all the allocated \acp{dpu} to be used.

Finally, the system proves to be beneficial for very large networks or large batch sizes, as it can take full advantage of the available memory. While \ac{wram} could potentially speed up the inference process, this is only effective when there is sufficient data reuse within the \ac{dpu}. Additionally, we observed that the majority of the time is spent on data transfers between the host and the device, which should not exist on real \ac{pim} devices, highlighting the need for intelligent memory controllers and cache coherence mechanisms. The comparison performed reached within the same order of magnitude as low-power \acp{gpu}. Overall, this work provides a clear view of what the system is capable of in these early stages. 

\section*{Acknowledgments}
This work was supported by Instituto de Telecomunica\c{c}\~{o}es and Funda\c{c}\~{a}o para a Ci\^{e}ncia e a Tecnologia (FCT), under projects UIDB/EEA/50008/2020 (DOI: 10.54499/UIDB/50008/2020), 2022.06780.PTDC, LA/P/0109/2020 (DOI: 0.54499/LA/P/0109/2020), and Ph.D. scholarship 2020.07124.BD. This work was also supported by the International Iberian Nanotechnology Laboratory, by the Digital Europe Programme under Grant  Agreement 101083770, and by the Recovery and Resilience Plan under the  European Union’s (EU) Recovery and Resilience Facility (RRF), framed  within the Next Generation EU, for the period 2021--2026, as part of the  ATTRACT project, with reference 774. The authors would also like to thank UPMEM for providing the servers used to run the experiments.

\bibliographystyle{unsrt}  
\bibliography{references}

\end{document}